# Enhanced self phase modulation in silicon nitride waveguides with integrated 2D MoS$_2$ films

*Shahaz S. Hameed, Di Jin, Aihao Zhao, Jiayang Wu \*, Junkai Hu, Irfan H. Abidi \*, Sebastien Cueff, Christian Grillet, Yuning Zhang, Houssein El Dirani, Corrado Sciancalepore, Sebastien Kerdiles, Quentin Wilmart, Sumeet Walia \*, Christelle Monat \*, and David J. Moss \**

S. Hameed, D. Jin, A. Zhao, J. Wu, J. Hu, D. J. Moss
Optical Sciences Centre
Swinburne University of Technology
Hawthorn, VIC 3122, Australia
E-mail: jiayangwu@swin.edu.au; dmoss@swin.edu.au

S. Hameed, C. Monat, S. Cueff, C. Grillet
INL, Institut des Nanotechnologies de Lyon UMR CNRS 5270, Ecole Centrale de Lyon, INSA Lyon, Universite Claude Bernard Lyon 1, CPE Lyon, 69130 Ecully, France
E-mail: christelle.monat@ec-lyon.fr

I. Abidi, S. Walia
School of Engineering, RMIT University, Melbourne, VIC 3000, Australia
E-mail: Irfan.haider.abidi@rmit.edu.au; sumeet.walia@rmit.edu.au

Y. Zhang
School of Physics, Peking University, Beijing, 100871, China

H. E. Dirani [+], C. Sciancalepore [++], S. Kerdiles, Q. Wilmart
CEA-LETI, Université Grenoble Alpes, 38054 Grenoble, France.

D. Jin, J. Wu, J. Hu, D. J. Moss, I. Abidi, S. Walia
ARC Centre of Excellence in Optical Microcombs for Breakthrough Science (COMBS), Melbourne, Victoria 3000, Australia

[+] Present address: LIGENTEC SA, Corbeil-Essonnes 91100, France

[++] Present address: STMicroelectronics, Grenoble, 38019, France.

## Abstract

On-chip integration of 2D materials provides a promising route towards next-generation integrated optical devices with performance beyond existing limits. Here, significantly enhanced spectral broadening induced by self-phase modulation (SPM) is experimentally demonstrated in silicon nitride ($Si_3N_4$) waveguides integrated with 2D monolayer molybdenum disulfide ($MoS_2$) films. Monolayer $MoS_2$ films with ultrahigh optical nonlinearity are synthesized via low-pressure chemical vapor deposition (LPCVD) and subsequently transferred onto $Si_3N_4$ waveguides, with precise control of the film coating length and placement achieved by selectively opening windows on the chip silica upper cladding. Detailed SPM measurements at telecom wavelengths are performed using fabricated waveguides with various $MoS_2$ film coating lengths. Compared to devices without $MoS_2$, increased spectral broadening of sub-picosecond optical pulses is observed for the hybrid devices, achieving a broadening factor of up to ~2.4 for a device with a 1.4-mm-long $MoS_2$ film. Theoretical fitting of the experimental results further reveals an increase of up to ~27 fold in the nonlinear parameter ($\gamma$) for the hybrid $MoS_2$ / $Si_3N_4$ waveguides and an equivalent Kerr coefficient ($n_2$) of $MoS_2$ nearly 5 orders of magnitude higher than $Si_3N_4$. These results confirm the effectiveness of on-chip integration of 2D $MoS_2$ films to enhance the nonlinear optical performance of integrated photonic devices.

# 1. Introduction

As a fundamental third-order ($\chi^{(3)}$) nonlinear optical process, self-phase modulation (SPM) occurs when high-peak-power optical pulses propagate through a nonlinear medium, where the nonlinear optical Kerr effect induces an intensity-dependent refractive index change that results in phase modulation and pulse spectral broadening [1-3]. As a key all-optical modulation mechanism, SPM has found wide applications in realizing broadband optical sources [4, 5], optical spectroscopy [6, 7], optical logic operations [8, 9], pulse compression [10, 11], optical diodes [12, 13], optical modulators [14, 15], and optical coherence tomography [16, 17].

Realizing SPM based on photonic integrated circuits (PICs) offers attractive advantages such as compact device footprint, high stability and scalability, and cost-effective mass production [18-21]. Benefiting from excellent compatibility with well-developed complementary metal-oxide-semiconductor (CMOS) fabrication technology, silicon has been a dominant material platform for PICs [22, 23]. Despite having a Kerr coefficient $n_2$ over an order of magnitude higher than silicon nitride ($Si_3N_4$) [24], the strong two-photon absorption (TPA) of silicon at near-infrared wavelengths, along with a resulting free carrier penalty, pose a fundamental limitation for its nonlinear optical performance in the telecom band [25]. As alternatives, other CMOS-compatible material platforms such as $Si_3N_4$ and high-index doped silica glass (Hydex), which exhibit negligible TPA in the telecom band, have been widely studied [26, 27]. Nevertheless, their relatively low Kerr nonlinearities ($n_2 = \sim 2.6 \times 10^{-19}$ $m^2$/W for $Si_3N_4$ and $n_2 = \sim 1.3 \times 10^{-19}$ $m^2$/W for Hydex, both over an order of magnitude lower than silicon [28, 29]) remain a limiting factor for the efficiency of nonlinear optical processes in these devices [30, 31].

To address the limitations of existing photonic integrated platforms, the incorporation of two-dimensional (2D) materials with ultrahigh optical nonlinearity has emerged as a promising strategy. The remarkable Kerr nonlinearity of 2D materials such as graphene [31-35], graphene oxide (GO) [36-40], molybdenum disulfide ($MoS_2$) [41-45], and tungsten disulfide ($WS_2$) [46] has been extensively studied and leveraged to implement nonlinear optical devices that overcome performance barriers and offer new capabilities.

Recently, we reported the synthesis of high-quality 2D $MoS_2$ films with precise control over defect formation [47, 48]. In this work, we integrate as-prepared 2D $MoS_2$ films onto $Si_3N_4$ waveguides and experimentally demonstrate significantly improved SPM performance for the hybrid $MoS_2$ / $Si_3N_4$ devices. High-quality monolayer $MoS_2$ films are synthesized via low-pressure chemical vapor deposition (LPCVD) and integrated onto $Si_3N_4$ waveguides through a polymer-assisted transfer process [49, 50]. Window opening in the silica upper cladding of the $Si_3N_4$ waveguides prior to this transfer enables precise control over the film coating length and placement on photonic integrated chips. We performed detailed SPM measurements at telecom wavelengths (~1550 nm) for hybrid waveguides with various $MoS_2$ film coating lengths and under increasing pulse peak powers up to ~91 W. Results show that the hybrid waveguides exhibit larger spectral broadening of sub-picosecond optical pulses compared with bare $Si_3N_4$ waveguides, achieving a maximum broadening factor (*BF*) of ~2.4. By fitting the experimental results with theory, we obtain an improvement of up to ~27 times in the nonlinear parameter (*γ*) for the hybrid waveguides, as well as an equivalent Kerr coefficient ($n_2$) of $MoS_2$ that is about 5 orders of magnitude higher than $Si_3N_4$. These results highlight the strong potential of on-chip integration of 2D $MoS_2$ films for enhancing the nonlinear optical performance of $Si_3N_4$ devices at telecom wavelengths.

## 2. Device design and fabrication

**Figure 1(a)** illustrates the atomic structure of a MoS$_2$ monolayer, where a hexagonal sheet of molybdenum (Mo) atoms is sandwiched between two hexagonal sheets of sulfur (S) atoms, with adjacent Mo and S atoms linked by strong covalent bonds. As a typical transition metal dichalcogenide [40, 51], MoS$_2$ is a semiconducting material with a direct bandgap of ~1.8–1.9 eV in its monolayer form, along with moderate carrier mobility, atomic-scale thickness (~0.7 nm), and high stability. These properties have enabled a wide range of photonic, electronic, and optoelectronic applications, such as nonlinear optical devices [41, 42, 52], field-effect transistors [53-55], optical modulators [56-58], and broadband photodetectors [59-61]. It is also worth noting that the quality of 2D MoS$_2$ films, which is strongly affected by intrinsic structural defects such as sulfur vacancies [42, 43], plays a critical role in determining their optical properties such as optical bandgap, refractive index, light absorption, and optical nonlinearity.

**Figure 1(b)** illustrates a Si$_3$N$_4$ waveguide integrated with a monolayer MoS$_2$ film. The cross-section view and top view of this hybrid waveguide is shown in **Figure 1(c)** and **Figure 1(d)**, respectively. In this work, we choose Si$_3$N$_4$ as the integrated platform material, which has a large bandgap of ~5.0 eV [62] that yields negligible TPA at near infrared wavelengths. A window is opened on the silica upper cladding to locally expose the Si$_3$N$_4$ waveguide to air, enabling MoS$_2$ film coating and its interaction with the waveguide evanescent field on a selected location.

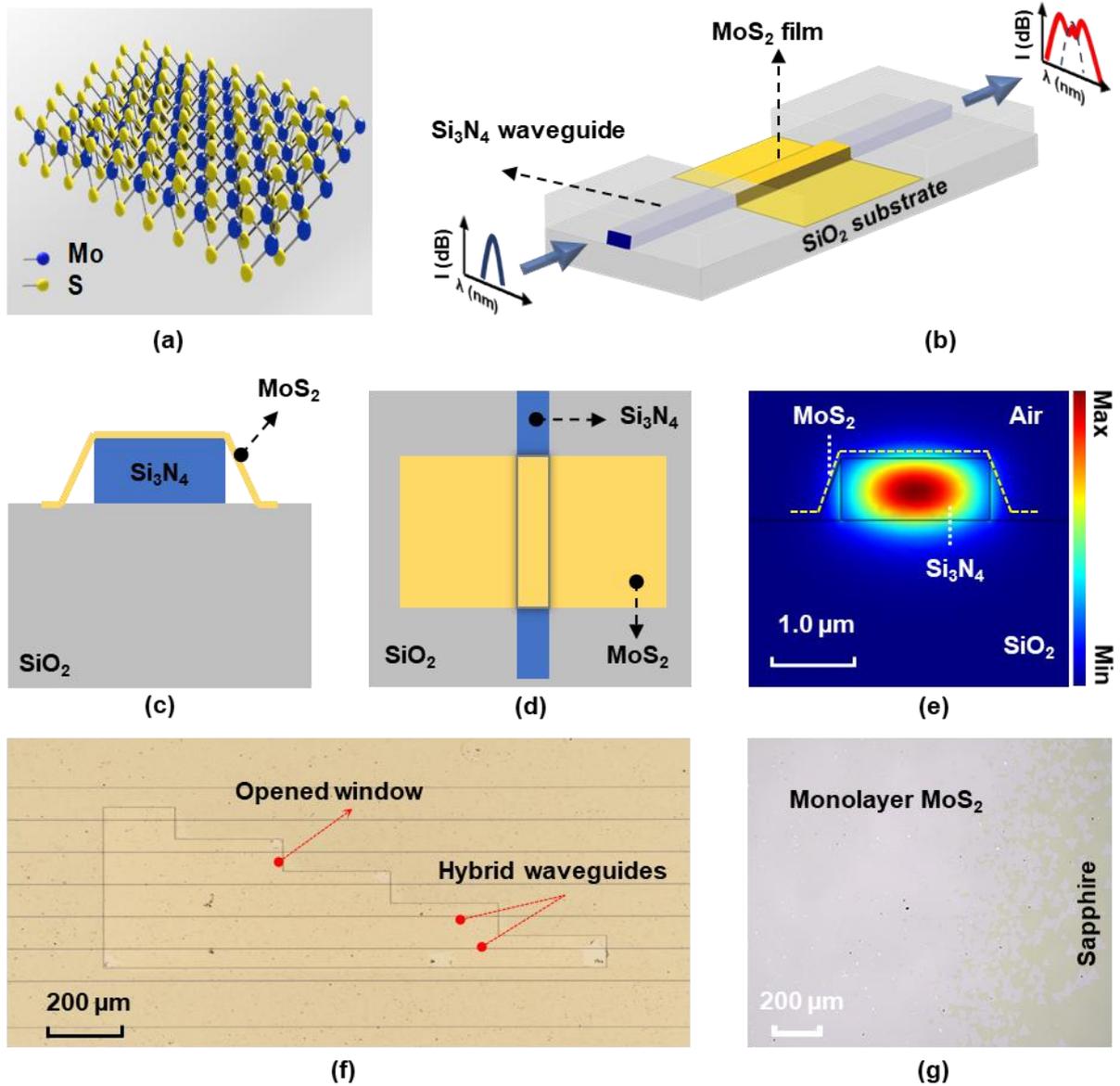

**Figure 1**. (a) Schematic atomic structure of monolayer molybdenum disulfide ($MoS_2$). (b) Schematic of a $Si_3N_4$ waveguide integrated with a monolayer $MoS_2$ film. The pulse schematics on the left and right sides illustrate the optical pulse spectra before and after propagation through the waveguide, respectively, exhibiting the SPM effect. (c) – (d) Schematic cross-section view and top view of the hybrid waveguide in (b), respectively. (e) Fundamental TE mode profile of the hybrid waveguide in (b). (f) Optical microscope image of a window-opened $Si_3N_4$ chip coated with monolayer $MoS_2$. (g) Optical microscope image of a monolayer $MoS_2$ film grown on a sapphire substrate.

**Figure 1(e)** shows the fundamental transverse electric (TE) mode profile (at 1550 nm) for the hybrid waveguide, which was simulated using commercial mode-solving software (COMSOL Multiphysics). In our simulation, the width and height of the $Si_3N_4$ waveguide were $W = 1.60$ μm and $H = 0.66$ μm, respectively. The thickness of the monolayer $MoS_2$ film was ~0.7 nm. The refractive index ($n$) and extinction coefficient ($k$) of $MoS_2$ for TE polarization measured at 1550 nm were $n_{TE}$ = ~3.8 and $k_{TE}$ = ~0.107, respectively. These values were obtained from our previous measurements, including both ellipsometry measurements of the complex refractive indices and results obtained by measuring hybrid integrated devices, as reported in Refs. [47, 49]. The light-matter interaction between the highly nonlinear $MoS_2$ film and the waveguide evanescent field can boost the nonlinear optical processes in the hybrid waveguide, forming the basis for enhancing the SPM performance. It is also worth mentioning that, due to the polymer-assisted transfer method employed in our device fabrication (which will be introduced later), the monolayer $MoS_2$ film does not conformally cover the sidewalls of the $Si_3N_4$ waveguide, leaving air gaps between the sidewalls and the film. This, however, has minimal influence on the $MoS_2$ mode overlap in the hybrid waveguide, as it mainly depends on the interaction between the evanescent field and the $MoS_2$ film on the waveguide top surface. Mode simulations further indicate that the mode overlap with the $MoS_2$ film near the sidewalls in **Figure 1e** (~0.0025%) is more than an order of magnitude lower than that on the top surface (~0.027%).

**Figure 1(f)** shows an optical microscope image of the fabricated $Si_3N_4$ chip coated with a monolayer $MoS_2$ film. In our fabrication, we first fabricated the bare $Si_3N_4$ waveguides using a CMOS-compatible, annealing-free, and crack-free method reported in Refs. [63, 64]. The process began with a two-step deposition of $Si_3N_4$ film (330 nm-thick layer in each step), achieved via LPCVD for strain management and crack prevention. Subsequently, 248-nm deep ultraviolet lithography (DUV) and fluorine-based dry etching were employed to pattern the waveguide layout. In detail, a $CF_4$-$CH_2F_2$-$O_2$ chemistry was used in a 300-mm reactor under 32 mTorr pressure and 150°C process temperature. An etch rate of 110 nm/minute was estimated, and the etching selectivity of the lithography resist against the $Si_3N_4$ layer was 1:1.5. After patterning the waveguide layout, a 2.2-µm-thick silica upper cladding was deposited via high-density plasma enhanced chemical vapor deposition (HDP-PECVD), followed by lithography and RIE to open windows in the silica cladding and etch down to the boxing layer. The RIE process employed $CHF_3$ / $CF_4$ / Ar gases, with the gas ratios appropriately adjusted to obtain an etch selectivity exceeding 5 for $SiO_2$ over $Si_3N_4$. The opened windows on the silica cladding allows the coating of the locally uncovered $Si_3N_4$ waveguides with $MoS_2$ films. All the fabricated $Si_3N_4$ waveguides had the same length of $L = \sim2.0$ cm. The lengths of the opened windows, which equal to the $MoS_2$ film coating lengths ($L_c$) for the hybrid waveguides, varied between ~0.2 mm and ~1.4 mm, enabling precise control of the $MoS_2$ film lengths that interacts with the $Si_3N_4$ waveguides.

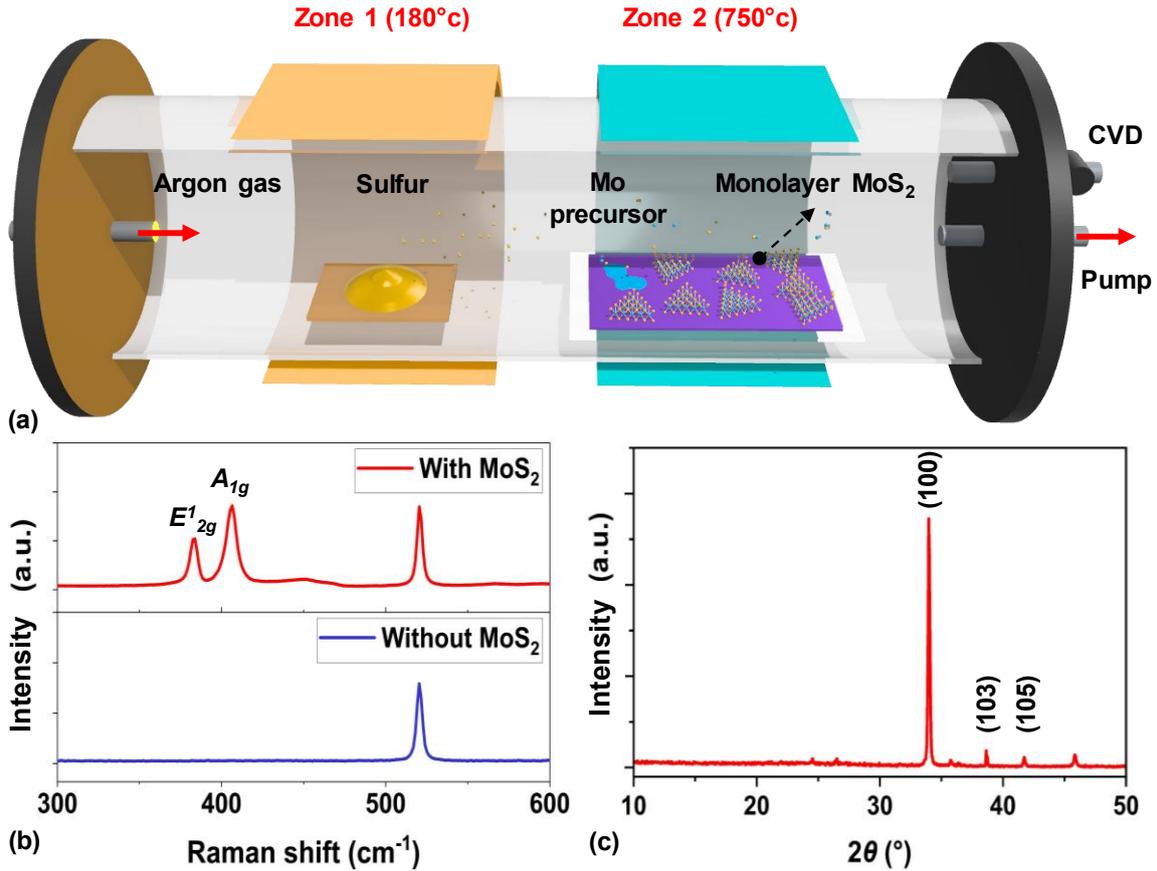

**Figure 2.** (a) Schematic illustration showing the chemical vapor deposition (CVD) process flow used for synthesizing monolayer $MoS_2$ films. (b) Measured Raman spectra of a $Si_3N_4$ chip before (bottom) and after coating a monolayer $MoS_2$ film (top). (c) X-ray diffraction (XRD) spectrum for the coated $MoS_2$ film on the $Si_3N_4$ chip.

Following the fabrication of the $Si_3N_4$ waveguides, a LPCVD method we developed in Ref. [47] was used to synthesize high-quality $MoS_2$ films on a sapphire substrate. This approach enables precise control of intrinsic atomic defects arising from sulfur vacancies, and facilitates the direct growth of large-area $MoS_2$ films with excellent uniformity. As illustrated in **Figure 2(a)**, $MoS_2$ monolayers were synthesized by a chemical vapor deposition (CVD) process carried out in a two-temperature-zone furnace. The process began by drop-casting a ~5-μL droplet of Mo precursor (an aqueous solution of ammonium molybdate tetrahydrate) onto an ultrasonically cleaned sapphire substrate placed in Zone 2, which was maintained at 750 °C. After that, the samples were allowed to dry at 110 °C for 5 minutes to remove residual solvent before loading into the CVD furnace. Subsequently, sulfur powder (~200 mg) heated to 180 °C in Zone 1 was vaporized and transported downstream into Zone 2 by a 70-sccm argon gas flow. Finally, the Mo and S species reacted on the substrate surface to yield $MoS_2$ single crystals. The growth for a relatively longer duration resulted in merging and thus continuous $MoS_2$ film growth. During the entire process, a chamber pressure of ~1 torr was sustained, thereby preserving the low-pressure CVD environment. The resulting $MoS_2$ film on sapphire is shown in **Figure 1(g)**.

The as-synthesized monolayer $MoS_2$ film was transferred onto the fabricated $Si_3N_4$ chip via a polymer-assisted transfer process [50], which is commonly used for on-chip transfer of 2D materials and is straightforward to implement. The as-grown film was first spin-coated with a polystyrene (PS) support layer, after which the PS/$MoS_2$ stack was exfoliated from the growth substrate via a water-assisted delamination process driven by interfacial surface energy differences [65]. The stack was then stamped onto the $Si_3N_4$ chip through van der Waals interactions, and finally, the PS layer was removed by dissolution in toluene.

**Figure 2(b)** shows the Raman spectra of the same $Si_3N_4$ chip before and after coating a monolayer $MoS_2$ film, which were characterized using a ~514 nm excitation laser with a spot size of ~1 μm. Except for a peak at ~517 cm$^{-1}$ arising from the substrate, two additional peaks at ~384 cm$^{-1}$ and ~404 cm$^{-1}$ were observed in the Raman spectrum for the $MoS_2$-coated chip, which corresponds to the in-plane ($E^1_{2g}$) and out-of-plane ($A_{1g}$) vibrational modes of $MoS_2$, respectively. The full widths at half maximum (FWHMs) of the $E^1_{2g}$ and $A_{1g}$ modes are ~3.17 cm$^{-1}$ and ~6.82 cm$^{-1}$, respectively. Moreover, the frequency separation of ~20 cm$^{-1}$ confirms the monolayer nature of the $MoS_2$ film. These features are consistent with previously reported CVD-grown monolayer $MoS_2$ [42, 56, 57], providing evidence for successful integration of high-quality 2D $MoS_2$ film onto the $Si_3N_4$ chip. This is also confirmed by the X-ray diffraction (XRD) spectrum for the coated MoS$_2$ film shown in **Figure 2(c)**. The diffraction peaks can be indexed to the (100), (103), and (105) reflections of hexagonal MoS$_2$ crystal structure, showing good agreement with the XRD spectra for MoS$_2$ reported in Refs. [58, 59]. Previously, we also performed many other characterizations for the as-synthesized MoS$_2$ films, as reported in Refs. [47, 49, 66, 67]. Due to the lack of access to a suitable slicing machine, we did not cleave the hybrid chip to observe the waveguide cross section. In fact, the MoS$_2$ film around the waveguide sidewalls, even if partially attached, does not significantly affect the TE mode profile studied in this work, as the mode interaction is dominated by the mode overlap with the MoS$_2$ film on the waveguide top surface.

## 3. Loss measurements

The integration of MoS$_2$ films onto Si$_3$N$_4$ waveguides introduces additional propagation loss. **Figure 3** shows a schematic of the experimental setup used for both loss measurements in this section and SPM measurements in **Section 4**. In our measurements, lensed fibers were employed to butt couple light into and out of the fabricated devices with inverse-taper couplers at both ends, in which the waveguide core width gradually narrows towards the chip facet. The taper tip width and length were ~120 nm and ~300 μm, respectively. The fiber-to-chip coupling loss was ~7.5 dB / facet. This coupling loss is relatively high because we did not use lensed fibers with very high numerical aperture (NA). In addition, a larger air gap was intentionally maintained between the lensed fiber and the chip to minimize heat transfer and mechanical vibrations. With optimized coupling conditions, the coupling loss can be reduced to ~ 4–5 dB /facet. For both the loss and SPM measurements, TE-polarized input light was selected, as it enables in-plane interaction between the waveguide evanescent field and the MoS$_2$ film, in contrast to the much weaker out-of-plane interaction due to the significant anisotropy of 2D materials [31, 36, 63].

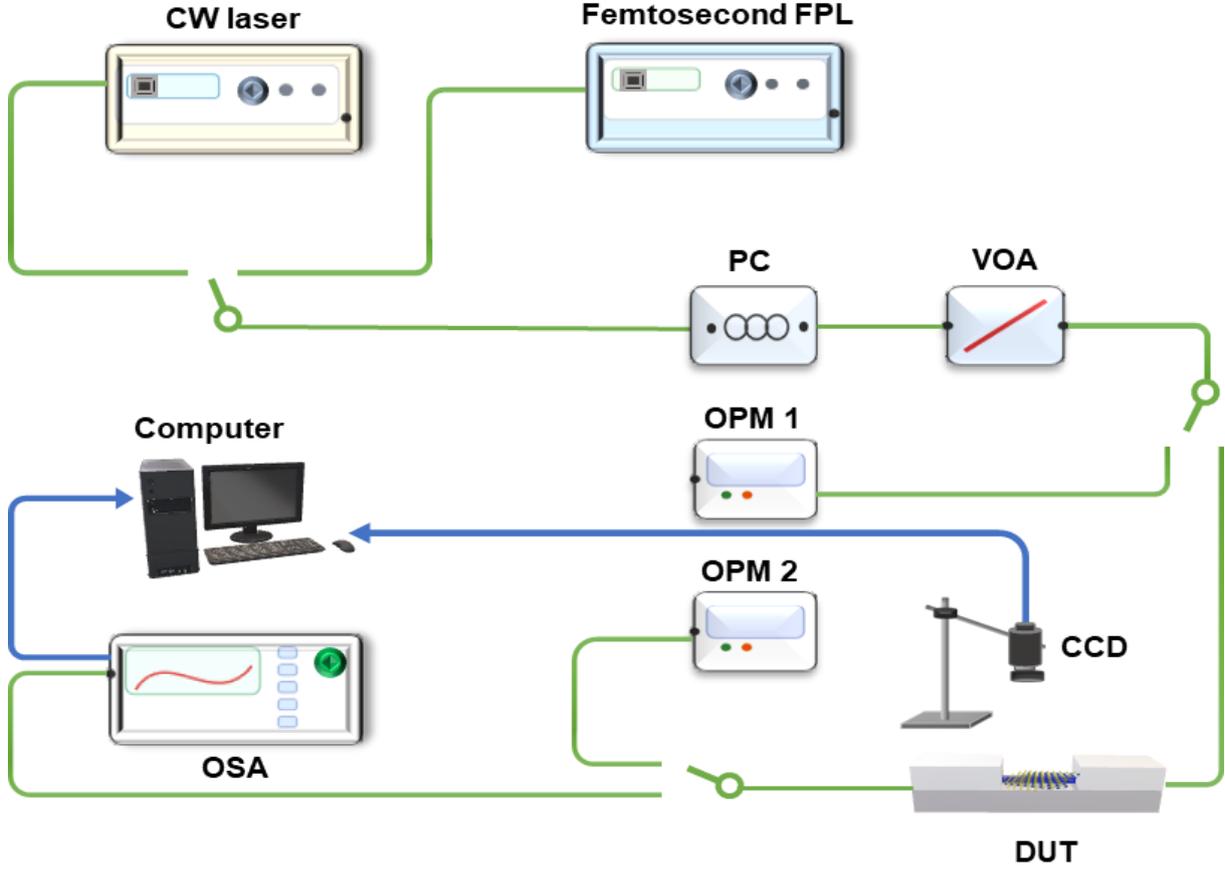

**Figure 3**. Experimental setup used for measuring loss and SPM for $MoS_2$-coated $Si_3N_4$ waveguides. CW laser: continuous-wave laser. FPL: fiber pulsed laser. PC: polarization controller. VOA: variable optical attenuator. OPM: optical power meter. DUT: device under test. CCD: charge-coupled device. OSA: optical spectrum analyzer.

We first measured the linear loss of the fabricated devices using a continuous-wave (CW) light. **Figure 4(a)** shows the measured TE-polarized insertion loss (IL) versus $MoS_2$ film coating length $L_c$ for the hybrid waveguides. For comparison, the corresponding results for the uncoated $Si_3N_4$ waveguides (*i.e.*, $L_c = 0$ mm) are also shown. For the uncoated waveguides with window lengths ranging from ~0.2 mm to ~1.4 mm, the IL variation was less than 0.2 dB. For the plot shown in **Figure 4(a)**, the uncoated waveguides with a window length of ~1.4 mm were employed. For different devices, the power and wavelength of the input CW light were kept the same as $P_{CW}$ = ~-10 dBm and $\lambda$ = ~1550 nm, respectively. Unless otherwise specified, the input power of CW light or optical pulses in our following discussion indicates the power coupled into the waveguide after excluding the input fiber-to-chip coupling loss (*i.e.*, ~7.5 dB). The data points depict the average values of measurements on three duplicate devices, and the error bars reflect the variations among them. As can be seen, the IL increases almost linearly with $L_c$. By using the cut-back method [68], we obtained a propagation loss of ~0.5 dB/cm for the $Si_3N_4$ waveguides with silica upper cladding. For the window-opened area (uncoated with $MoS_2$), the propagation loss is ~1.5 dB/cm. This value was obtained by fitting the transmission spectra of microring resonators with opened windows on the same chip using the scattering matrix method [69, 70].

Based on these values and the measured IL values in **Figure 4(a)**, the excess propagation loss induced by monolayer $MoS_2$ film was extracted to be ~24 dB/cm. Compared with our earlier reports on $Si_3N_4$ waveguides coated with monolayer GO films [71, 72], the loss induced by $MoS_2$ is about 10 times higher, yet remains about five times lower than graphene-induced excess propagation loss in graphene-coated $Si_3N_4$ waveguides [35, 63, 73]. The extinction coefficient of $MoS_2$ extracted from the measured excess propagation loss is $k$ = ~0.107 (as used for the simulation in **Figure 1(e)**), which is typical for $MoS_2$ and shows agreement with the value obtained from our previous measurements using silicon devices [49]. The extracted $k$ of monolayer $MoS_2$ here is about one order of magnitude higher than that of monolayer GO (i.e., $k \approx 0.01$) [74]. We also notice that 2D $MoS_2$ films with significantly lower loss have been reported in Ref. [75]. We therefore infer that the higher loss of our synthesized $MoS_2$ films may be attributed to factors such as the presence of defects or residual metal ions introduced during the CVD process. Optimization of the loss of our $MoS_2$ films will be the subject of future work.

**Figure 4(b)** shows the measured TE-polarized IL versus input CW power for an uncoated $Si_3N_4$ waveguide and the hybrid waveguide with a 1.4-mm-long $MoS_2$ film. In our following discussion, the comparison between uncoated $Si_3N_4$ waveguides and hybrid waveguides refers to the same devices (*i.e.*, with opened windows) before and after coating with $MoS_2$ films. For both waveguides, the IL does not show any obvious variation with the input CW power. This indicates that the $MoS_2$ film exhibited negligible power-dependent loss in the measured power range. At an input CW power of ~400 mW (*i.e.*, ~2 W prior to waveguide coupling), we had already reached the maximum output of our erbium-doped fiber amplifier. This amplifier is not shown in **Figure 3** since it was only employed in the experiments for **Figure 4b**. Yet, no significant changes were observed in the measured IL for the hybrid waveguide, even after repeating the entire power-tuning process several times (the slight variations among these are reflected by the error bars in **Figure 4b**). This indicates the excellent thermal stability of $MoS_2$, a key advantage for nonlinear optical applications that typically require high powers to boost nonlinear interactions.

**Figure 4(c)** shows the measured TE-polarized IL versus input CW wavelength for the two waveguides measured in **Figure 4(b)**. The input CW power was maintained at ~-10 dBm. Due to the limited wavelength tuning range for our CW laser, we could only measure the IL between 1500 and 1600 nm. Within this range, the flat spectral response of both uncoated and $MoS_2$-coated $Si_3N_4$ waveguides indicates negligible wavelength-dependent material absorption or coupling loss.

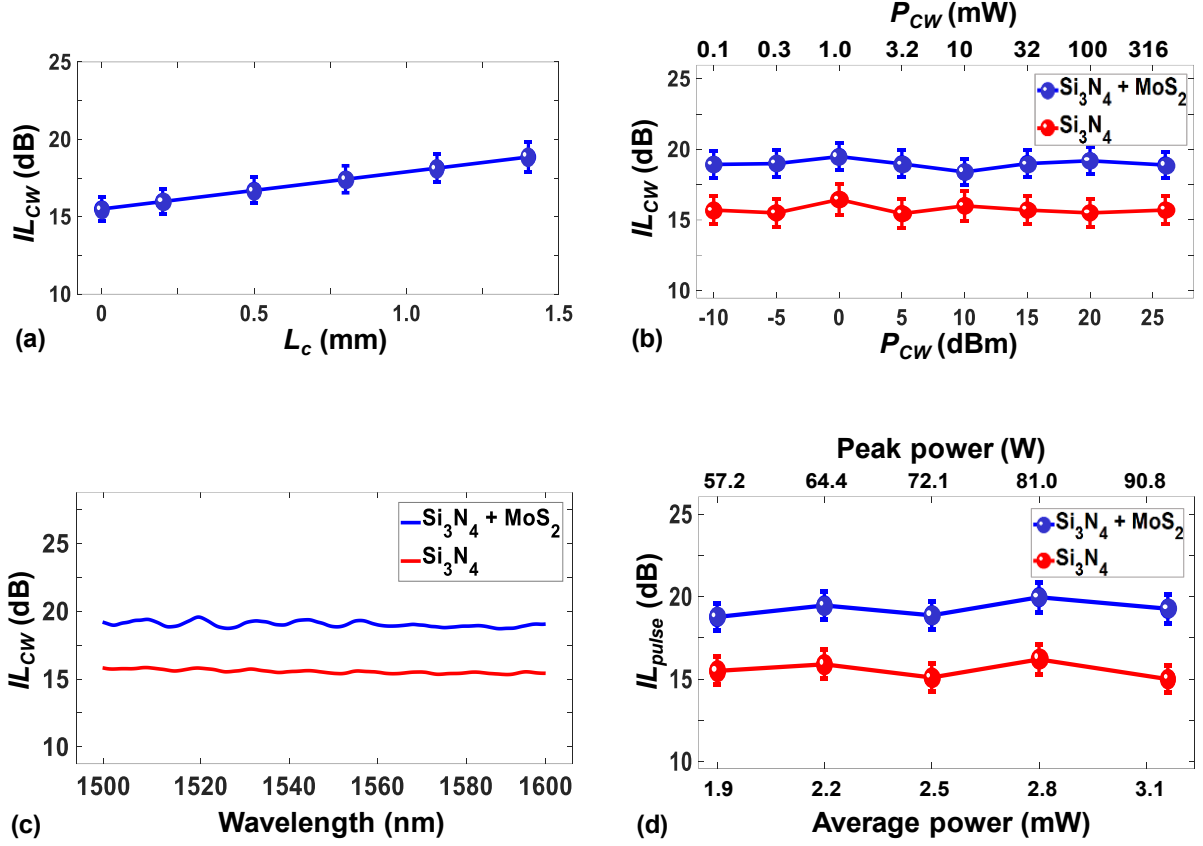

**Figure 4.** (a) Measured insertion loss ($IL_{CW}$) of hybrid MoS$_2$-Si$_3$N$_4$ waveguides versus MoS$_2$ coating length ($L_c$). The corresponding results for uncoated Si$_3$N$_4$ waveguides (*i.e.*, $L_c$ = 0 mm) are also shown for comparison. (b) Measured insertion loss ($IL_{CW}$) versus input CW power ($P_{CW}$) for an uncoated Si$_3$N$_4$ waveguide and a hybrid waveguide with a 1.4-mm-long MoS$_2$ film. (c) Measured insertion loss ($IL_{CW}$) versus input CW wavelength for the two waveguides in (b). (d) Measured insertion loss ($IL_{pulse}$) versus input power of optical pulses for the two waveguides in (b). In (a) and (b), the input CW wavelength is ~1550 nm. In (a) and (c), the input CW power is ~ -10 dBm. In (a), (b), and (d), the data points depict the average of measurements on three samples, and the error bars illustrate the variations among the different samples.

In addition to linear loss measurements performed using CW light, we employed optical pulses generated by a fiber pulsed laser (FPL, as illustrated in **Figure 3**) to characterize the nonlinear loss of the fabricated devices. **Figure 4(d)** shows the measured IL versus input power of optical pulses for the two waveguides measured in **Figures 4(b)** and **4(c)**. The data points depict the average values of three measurements on the same devices, and the error bars reflect the variations among them. The FPLs generated nearly Fourier-transform limited optical pulses centered at ~1557 nm. The pulse duration and repetition rate were ~580 fs and ~60 MHz, respectively. The average input powers ranged between ~1.9 mW and ~3.2 mW, which corresponds to peak powers from ~57 W to ~91 W (*i.e.*, from ~321 W to ~512 W before the chip). In contrast to CW light, optical pulses have peak powers that are significantly higher than their average powers, making them suitable for characterizing nonlinear optical loss without introducing significant thermo-optic effects. Similar to those observed in **Figure 4(b)**, the measured IL of both uncoated and hybrid $Si_3N_4$ waveguides showed no significant variation with increasing input power of optical pulses. This indicates the absence of significant nonlinear optical absorption such as TPA and saturable absorption effects for both $Si_3N_4$ and $MoS_2$ – a behavior consistent with their large optical bandgaps (which exceed twice the photon energy at 1550 nm, *i.e.*, ~1.6 eV) as previously discussed.

## 4. SPM measurements

Following the loss measurements, we performed SPM measurements to characterize the spectral broadening of optical pulses after propagating through the fabricated devices. In our SPM measurements, we employed the same FPL and fabricated devices as those used for loss measurements in **Section 3**. As illustrated in **Figure 3**, optical pulses generated from the FPL were coupled into the fabricated devices, with a variable optical attenuator (VOA) and a polarization controller (PC) employed to adjust the power and polarization of the input light, respectively. After passing through the fabricated devices, the output signal was collected by an optical spectrum analyzer (OSA) to monitor spectral broadening. We did not employ any fiber couplers in our SPM measurements to avoid any spectral filtering distortions induced by them.

**Figure 5(a)** shows the normalized spectra of optical pulses before and after propagating through the uncoated $Si_3N_4$ waveguide and the hybrid waveguide with a 1.4-mm-long $MoS_2$ film. The peak power of the input sub-picosecond pulse was maintained at ~ 91 W. Compared with the input pulse spectrum, slight spectral broadening was observed in the output spectrum from the uncoated $Si_3N_4$ waveguide, which was mainly caused by SPM in the $Si_3N_4$ waveguide. The $MoS_2$-coated $Si_3N_4$ waveguide exhibited much greater spectral broadening than the uncoated $Si_3N_4$ waveguide despite the additional loss caused by the 1.4-mm $MoS_2$ layer, indicating the significantly enhanced SPM in the hybrid waveguide.

**Figure 5(b)** shows the normalized spectra of optical pulses after propagating through hybrid waveguides with various $MoS_2$ coating lengths ($L_c$). Here we measured five hybrid devices, with $L_c$ varying between ~0.2 mm and ~1.4 mm. The corresponding result for an uncoated $Si_3N_4$ waveguide is also shown for comparison. The uncoated waveguides with window lengths ranging from ~0.2 mm to ~1.4 mm showed no significant variation in their output spectra, so we plot the result for the waveguide with a window length of ~1.4 mm. It can be seen that more significant spectral broadening was observed for an increased $MoS_2$ coating length, further confirming that the enhanced SPM in the hybrid waveguides was induced by the coated $MoS_2$ films.

**Figure 5(c)** shows the normalized optical spectra measured at different input pulse peak powers ($P_{peak}$) for the hybrid waveguide with $L_c$ = 1.4 mm. Here we show the results for five different $P_{peak}$ values ranging from ~57 W to ~91 W – identical to those used for nonlinear loss measurements in **Figure 4(d)**. As the input peak power was raised, the output spectra exhibited increasing broadening, in agreement with the expected behavior of SPM-induced spectral broadening.

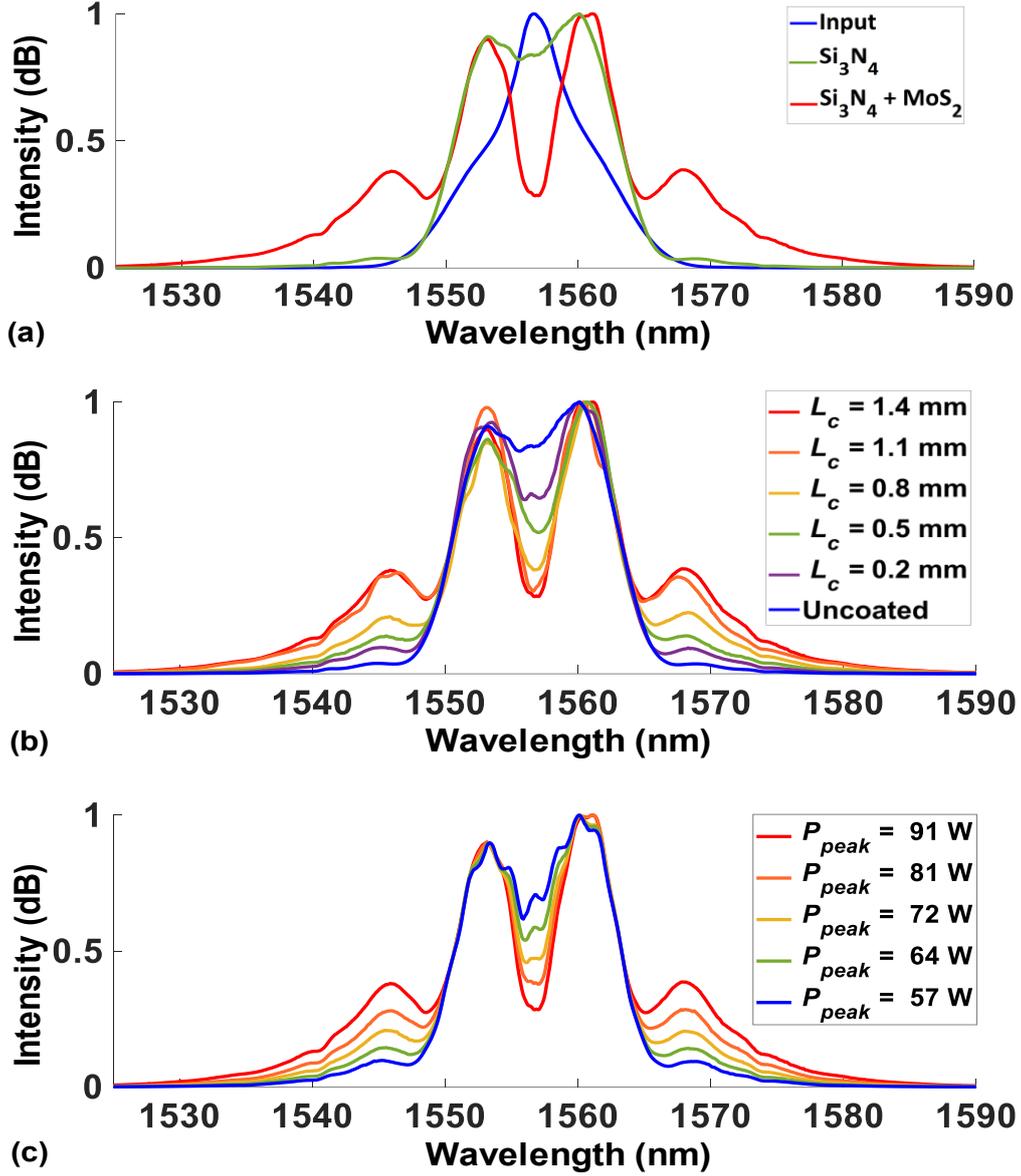

**Figure 5**. SPM experimental results. (a) Normalized spectra of optical pulses before and after propagating through an uncoated $Si_3N_4$ waveguide and a hybrid waveguide with a 1.4-mm-long $MoS_2$ film. (b) Normalized spectra of optical pulses after propagating through hybrid waveguides with various $MoS_2$ coating lengths ($L_c$) ranging between ~0.2 mm and ~1.4 mm. The output spectrum from an uncoated $Si_3N_4$ waveguide is also shown for comparison. (c) Normalized optical spectra measured at different input pulse peak powers ($P_{peak}$) for the hybrid waveguide with a 1.4-mm-long $MoS_2$ film. In (a) and (b), the $P_{peak}$ was kept the same as ~91 W.

## 5. Theoretical analysis and discussion

In this section, we analyze the SPM experimental results in **Section 4** by fitting them with theory to extract the nonlinear parameter $\gamma$ of the hybrid waveguide and the equivalent Kerr coefficient $n_2$ of $MoS_2$.

To quantitatively compare spectral broadening across different waveguides, broadening factors (*BF*) were calculated from the measured output spectra in **Figure 5**. The *BF* is defined as [33, 38, 63]:

$$BF = \frac{\Delta\omega_{rms}}{\Delta\omega_0} \quad (1)$$

where $\Delta\omega_{rms}$ and $\Delta\omega_0$ are the root-mean-square (RMS) spectral widths of the signals before and after propagation through the waveguides (either uncoated or coated with $MoS_2$), respectively. Note that we use the RMS spectral width here instead of the FWHM because spectra generated by SPM are generally non-Gaussian and may exhibit asymmetry or multiple peaks. Since the FWHM captures only the central part of the spectrum, the RMS spectral width, defined by the second-order moment of the spectral power distribution, provides a more appropriate measure of SPM-induced spectral broadening.

**Figure 6(a)** shows the calculated *BF*s for the optical spectra in **Figure 5(b)** measured at $P_{peak}$ = ~91 W. Increasing the $MoS_2$ coating length leads to a rise in the *BF*, showing a trend consistent with that in **Figure 5(b)**. At $L_c$ = ~1.4 mm, a maximum *BF* of ~2.4 is achieved for the hybrid waveguide. In comparison, the calculated *BF* for the uncoated $Si_3N_4$ waveguide is only ~1.2. **Figure 6(b)** shows the calculated *BF*s for the optical spectra measured from the hybrid waveguide with $L_c$ =1.4 mm, as was shown in **Figure 5(c)**. With the input pulse peak power increasing from ~57 W to ~91 W, the *BF* increases from ~1.4 to ~2.4.

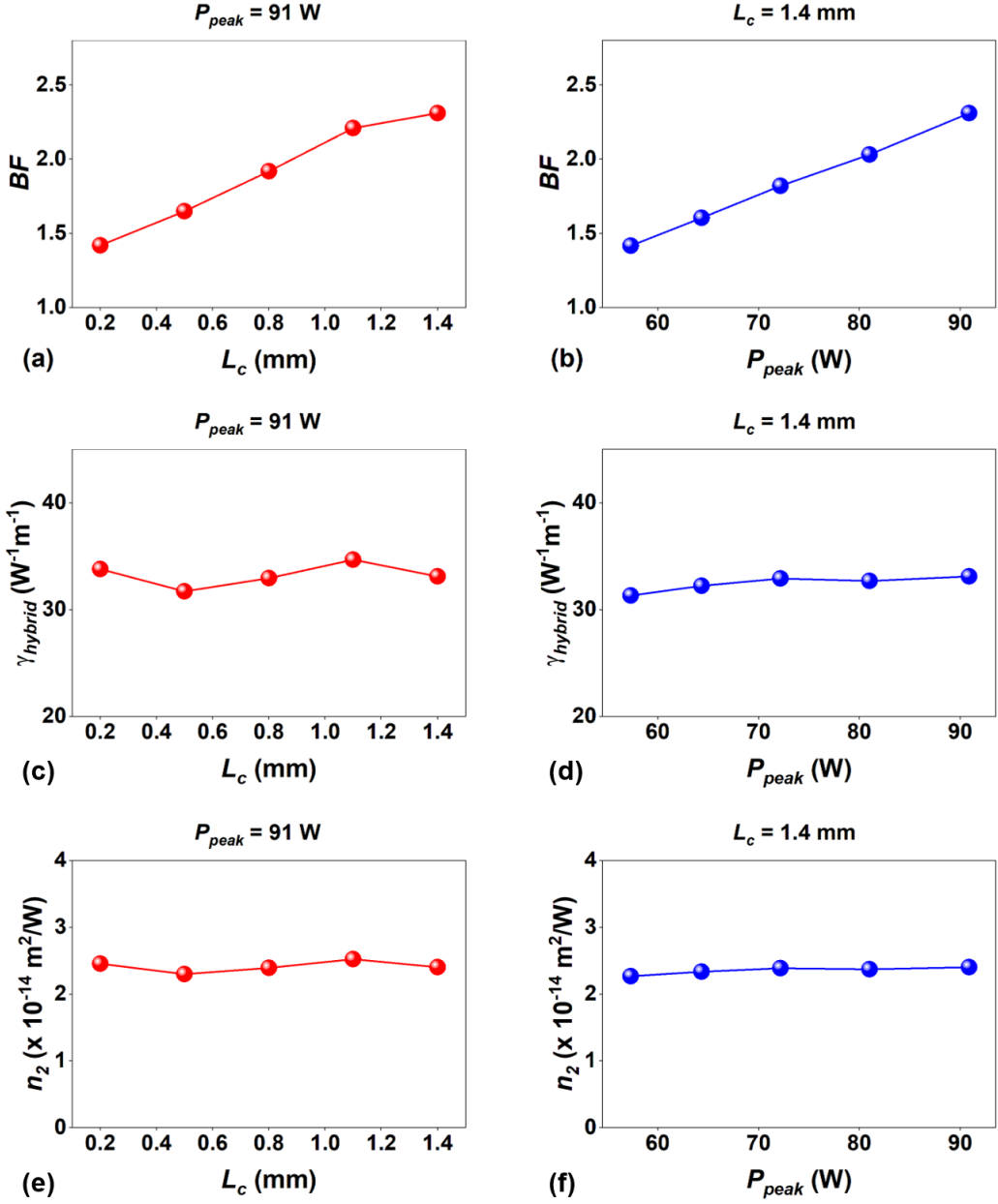

**Figure 6**. (a) Calculated broadening factor (*BF*) versus MoS$_2$ coating length ($L_c$) for the measured optical spectra in Figure 5(b). (b) Calculated BF versus input pulse peak power ($P_{peak}$) for the measured optical spectra in Figure 5(c). (c) – (d) Nonlinear parameters for the MoS$_2$-coated segments ($\gamma_{hybrid}$) in the hybrid waveguides obtained by fitting the results in Figures 5(b) and (c), respectively. (e) – (f) Equivalent Kerr coefficient ($n_2$) of MoS$_2$ extracted from the results in (c) and (d), respectively.

To fit the experimental results with theory, we simulated the SPM-induced spectral broadening of optical pulses by numerically solving the nonlinear Schrödinger equation as follows [21, 38]:

$$\frac{\partial A}{\partial z} = -\frac{i\beta_2}{2}\frac{\partial^2 A}{\partial t^2} + i\gamma |A|^2 A - \frac{1}{2}\alpha A \qquad (2)$$

where $A(z, t)$ is the slowly varying amplitude of the temporal pulse envelope along the propagation direction $z$, $\beta_2$ is the second-order dispersion coefficient related to the waveguide fundamental TE mode, $\gamma$ is the waveguide nonlinear parameter and $\alpha$ is the loss factor. In **Eq. (2)**, only the second-order dispersion term $\beta_2$ is considered, since the physical length of the waveguides (*i.e.*, ~2 cm) is much shorter than the dispersion length (> 1 m) [76]. The $\beta_2$ value used in our simulations ($-2.5 \times 10^{-26}$ s$^2$/m at 1550 nm) is consistent with that reported in our previous work for Si$_3$N$_4$ waveguides of the same geometry [62]. According to our measurements in **Figure 4(d)**, both Si$_3$N$_4$ and MoS$_2$ exhibited negligible nonlinear optical absorption, so only linear loss was considered for $\alpha$. In our simulation, we divided the MoS$_2$-coated Si$_3$N$_4$ waveguides into uncoated (with silica cladding) and hybrid sections (coated with MoS$_2$ films). For each section, **Equation (2)** was numerically solved, with the output of one section serving as the input for the next one. The output from the final section was used to fit the measured spectra in **Figure 5** and extract the waveguide nonlinear parameter $\gamma$.

We first extracted the $\gamma$ for the uncoated Si$_3$N$_4$ waveguide, which is ~1.2 W$^{-1}$m$^{-1}$ and consistent with previously reported values in Refs. [49, 62, 64]. Subsequently, we proceeded to extract the $\gamma$ values for the MoS$_2$-coated sections in the hybrid waveguides with various MoS$_2$ coating lengths. **Figures 6(c)** and **(d)** show the $\gamma$ values for the hybrid sections obtained by fitting the experimental results in **Figures 5(b)** and **(c)**, respectively. Despite variations in MoS$_2$ coating length and input pulse peak power, the extracted $\gamma$ values remain nearly unchanged, with variations of less than 3.6 W$^{-1}$m$^{-1}$. This reflects the consistency of our SPM measurements. The average $\gamma$ for the hybrid segments is ~32 ± 2 W$^{-1}$m$^{-1}$, which is about 27 times higher than that of the uncoated Si$_3$N$_4$ waveguide. This highlights the substantial improvement in the waveguide's optical nonlinearity after incorporating 2D MoS$_2$. We stress again that this optical nonlinear boost afforded by the hybrid section overcompensates for the additional loss induced there by the MoS$_2$, so that the overall SPM effect is enhanced at the output of the hybrid waveguide with respect to that of the uncoated one.

The equivalent Kerr coefficient ($n_2$) of 2D MoS$_2$ is further extracted based on the fit $\gamma$ values of the hybrid waveguides using [38, 77, 78]:

$$\gamma = \frac{2\pi}{\lambda_c} \frac{\iint_D n_0^2(x,y) n_2(x,y) |E(x,y)|^4 dxdy}{\left[\iint_D n_0(x,y) |E(x,y)|^2 dxdy\right]^2} \tag{3}$$

where $\lambda_c$ is the light wavelength, $D$ is the integral of the optical fields over the material regions, $E(x, y)$ is the electric field distribution calculated by the mode solving software, and $n_0 (x, y)$ and $n_2 (x, y)$ are the refractive index and Kerr coefficient profiles across the waveguide cross section, respectively. The values of $n_2$ for silica and Si$_3$N$_4$ used in our calculation were ~2.6 × 10$^{-20}$ m$^2$ / W [24] and ~2.5 × 10$^{-19}$ m$^2$ / W, respectively, with the latter obtained by fitting the $\gamma$ for the uncoated Si$_3$N$_4$ waveguide.

**Figures 6(e)** and **(f)** show the equivalent Kerr coefficient ($n_2$) of MoS$_2$ extracted from the fitted $\gamma$ values in **Figures 6(c)** and **(d)**, respectively. Consistent with the results in **Figures 6(c)** and **(d)**, the extracted $n_2$ does not show any significant variations with respect to the input power or MoS$_2$ coating length. The average equivalent $n_2$ of MoS$_2$ is ~(2.3 ± 0.2) × 10$^{-14}$ m$^2$ / W – about 5 orders of magnitude higher than Si$_3$N$_4$ at 1550nm. This highlights the ultrahigh third-order optical nonlinearity of monolayer MoS$_2$ as compared to bulk materials.

**Table 1. Comparison of integrated optical waveguides incorporating 2D materials for enhanced SPM. G: graphene. NLO: nonlinear optical.**

| 2D Material | Integrated WG [a] | $PL$ [b] (dB/mm) | 2D film thickness | $n_2$ of 2D material ($\times 10^{-14}$ m$^2$ W$^{-1}$) | NLO performance [c] | Ref. |
|---|---|---|---|---|---|---|
| GO | Si | 2.5 | 2–40 nm 1–20 layers | 1.2 – 1.4 | Max BF of 4.34 | [51] |
| GO | Si$_3$N$_4$ | 0.3 – 0.6 | 2 – 4 nm | 1.19 – 1.23 | $\gamma$ improvement from 1.5 to 27.6 W$^{-1}$m$^{-1}$ | [71] |
| G | Si | 200 | monolayer | — | $\gamma$ improvement from 150 to 510 W$^{-1}$m$^{-1}$ | [34] |
| WS$_2$ | Si$_3$N$_4$ | 2.5 | 5.6 nm | 0.20 – 0.22 | $\gamma$ improvement from 1.75 to 650 W$^{-1}$ m$^{-1}$ | [46] |
| MoS$_2$ | Si | — | 10 nm | 0.01 | Extracted $n_2$ of MoS$_2$ about 100 times that of Si | [45] |
| MoS$_2$ | Si$_3$N$_4$ | 2.4 | 0.7 nm | 2.1 – 2.5 | Extracted $n_2$ of MoS$_2$ about 5 orders higher than that of Si$_3$N$_4$ | This work |

[a] WG: waveguide.

[b] $PL$: the linear propagation loss of the hybrid waveguides with the lowest 2D film thickness.

[c] BF: broadening factor. γ: nonlinear parameter. The improvements in CE and γ are relative to the uncoated waveguides.

According to **Eq. (3)**, the nonlinear parameter $\gamma$ is determined by the weighted contributions of the Kerr coefficients ($n_2$) from different material regions in the waveguide cross section. Although the 2D MoS$_2$ film occupies only a small fraction, its exceptionally high Kerr coefficient enables a substantial enhancement of $\gamma$ for the hybrid waveguides. **Table 1** compares the performance for integrated optical waveguides incorporating different 2D materials for enhanced SPM. We note that the $n_2$ value of MoS$_2$ obtained here is much higher than those reported in Refs. [43, 45], which can probably be attributed to our film being strictly monolayer (in contrast to multi-layered films used in Refs. [43]), as well as its high quality with minimal defects enabled by our developed synthesis method [47].

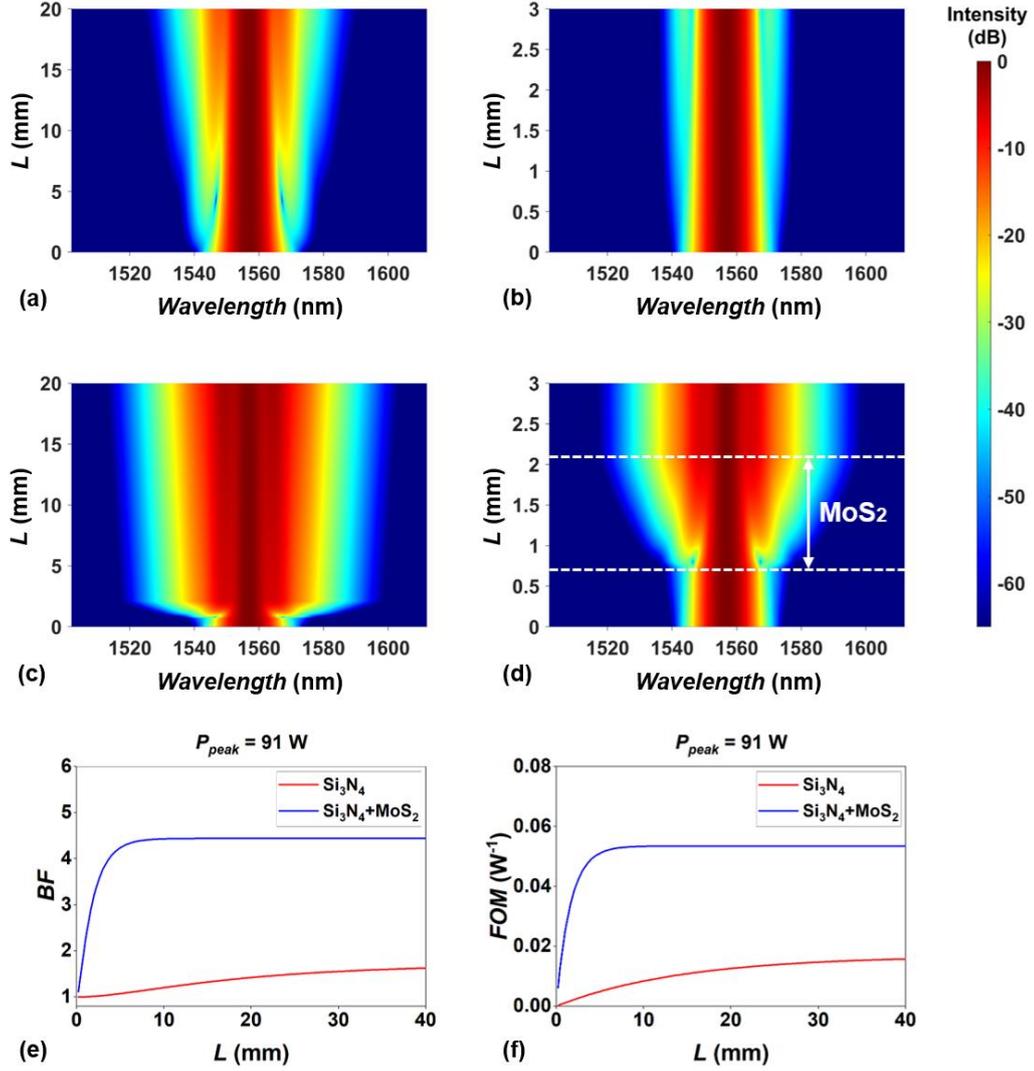

**Figure 7**. (a) Simulated spectral evolution of optical pulses propagating along an uncoated $Si_3N_4$ waveguide. (b) Zoom-in view of (a) around the window opening area ranging from 0.7 mm to 2.1 mm. (c) Simulated spectral evolution of optical pulses propagating along a hybrid waveguide coated with a 1.4-mm-long $MoS_2$ film. (d) Zoom-in view of (c) around the window opening area ranging from 0.7 mm to 2.1 mm. (e) Simulated *BF* versus waveguide length ($L$) for the uncoated $Si_3N_4$ waveguide and the hybrid waveguide coated with monolayer $MoS_2$ film. (f) *FOM* versus waveguide length ($L$) for the uncoated $Si_3N_4$ waveguide and the hybrid waveguide coated with monolayer $MoS_2$ film. In (a) – (d), the uncoated and hybrid waveguides are the same as those used for our SPM experiments. In (e) – (f), for comparison, the hybrid waveguide is assumed to be uniformly coated with $MoS_2$ (without uncoated segments). In (a) – (f), the input pulse peak power is ~91 W.

**Figures 7(a)** and **(c)** show the simulated spectral evolution along the entire length (*i.e.*, ~2 cm) of the uncoated and hybrid $Si_3N_4$ waveguides with $L_c$ = 1.4 mm, respectively. They were simulated by using the fitted $\gamma$ values for the uncoated and $MoS_2$-coated waveguides obtained in **Figure 6**. Zoom-in views around the window opening area (between 0.7 mm and 2.1 mm, near the input ports) are shown in **Figures 7(b)** and **(d)**, respectively. Unlike the gradual spectral broadening along the entire length of the uncoated $Si_3N_4$ waveguides, dramatical spectral broadening occurs within the $MoS_2$-coated section in the hybrid waveguide, followed by gradual broadening in the subsequent uncoated section. This further reflects enhanced SPM in the $MoS_2$-coated section enabled by integrating $MoS_2$ with ultrahigh optical nonlinearity.

To analyze the influence of waveguide length on the SPM performance, **Figure 7(e)** shows the simulated *BF* versus waveguide length (*L*) for the uncoated $Si_3N_4$ waveguide and the hybrid waveguide coated with monolayer $MoS_2$. To simplify the discussion, we assume that the hybrid waveguide is uniformly coated with $MoS_2$ (*i.e.*, $L = L_C$). As can be seen, the *BF* increases more rapidly when $L < 6$ mm and gradually saturates as $L$ increases further. This reflects the dependence of the *BF* for the output pulse on the waveguide length $L$, resulting from the trade-off between high optical nonlinearity and additional loss both introduced by the 2D $MoS_2$. Such trade-off is also reflected by the calculated figure of merit (*FOM*) for nonlinear optical waveguides in **Figure 7(f)**, which is a function of waveguide length $L$ defined by [37]:

$$FOM\,(L) = \gamma \cdot L_{eff}(L) \qquad (4)$$

where $\gamma$ is the waveguide nonlinear parameter and $L_{eff}(L) = [1 - exp(-\alpha \cdot L)]/\alpha$ is the effective interaction length, with $\alpha$ denoting the linear loss attenuation coefficient.

## 6. Conclusion

In this work, we experimentally demonstrate significantly enhanced SPM-induced spectral broadening of optical pulses in $Si_3N_4$ waveguides integrated with 2D $MoS_2$ films at telecom wavelengths. High-quality monolayer $MoS_2$ films are synthesized via LPCVD, followed by a polymer-assisted transfer process that enables their on-chip integration. Selective window openings in the upper waveguide cladding allow accurate control of the film coating length and position on the integrated chips. Detailed SPM measurements are carried out using fabricated devices with various $MoS_2$ film lengths and at various pulse peak powers. Compared with uncoated $Si_3N_4$ waveguides, enhanced spectral broadening of sub-picosecond optical pulses is observed in the output spectra of the hybrid waveguides, achieving a maximum *BF* of ~2.4. Analysis of the experimental results through theoretical fitting further reveals up to ~27-fold increase in the nonlinear parameter for the hybrid waveguides, together with an extracted equivalent $n_2$ of $MoS_2$ that is nearly five orders of magnitude higher than $Si_3N_4$. These results validate the integration of 2D $MoS_2$ films onto $Si_3N_4$ devices as a powerful approach to enhance their Kerr nonlinear optical performance at telecom wavelengths.

**Acknowledgements**

This work was supported by the Australian Research Council Centre of Excellence in Optical Microcombs for Breakthrough Science (Grant No. CE230100006), the Australian Research Council Discovery Projects Programs (Grant Nos. CE170100026, DP220100020, and DP240100145), the Agence Nationale de la Recherche (Grant No. MIRthFUL, ANR-21-CE24-0005), a France 2030 government grant (Grant No. ANR-22-PEEL-0005), the International Associated Laboratory in Photonics between France and Australia (LIA-ALPhFA), and the Australia France Network of Doctoral Excellence (AUFRANDE). The AUFRANsDE project has received funding from the European Union's Horizon Europe research and innovation program under the Marie Sklodowska-Curie HORIZON-MSCA-2021-COFUND-01 grant agreement (No. 101081465). The integrated photonic chip was manufactured at Cea-Leti facilities in Grenoble, France.

**Author contributions**

J.W. conceived the idea and designed the research. H.E.D., C.S., S.K., Q.W. designed and fabricated the $Si_3N_4$ waveguides. I.H.A performed $MoS_2$ film synthesis and on-chip transfer. D.J. and J.H. performed material and device characterization. S.S.H., D.J., and J.W. performed loss and SPM measurements. S.S.H. performed data processing and prepared the initial manuscript. A.Z. and Y.Z. performed theoretical simulations and fitting of experimental results. J.W., S.W., C.M., and D.M jointly supervised this project. S.C. and C.G. are on the joint supervision team. All authors participated in the review and discussion of the manuscript.

**Conflict of Interest**

The authors declare no competing financial interest.

**References**

[1] R. H. Stolen, and C. Lin, "Self-phase-modulation in silica optical fibers," *Physical Review A,* vol. 17, no. 4, pp. 1448-1453, 1978.

[2] G. P. Agrawal, and N. A. Olsson, "Self-phase modulation and spectral broadening of optical pulses in semiconductor laser amplifiers," *IEEE Journal of Quantum Electronics,* vol. 25, no. 11, pp. 2297-2306, 1989/11//, 1989.

[3] L. Wu, X. Yuan, D. Ma, Y. Zhang, W. Huang, Y. Ge, Y. Song, Y. Xiang, J. Li, and H. Zhang, "Recent Advances of Spatial Self-Phase Modulation in 2D Materials and Passive Photonic Device Applications," *Small,* vol. 16, no. 35, pp. 2002252, 2020, 2020.

[4] H. Guo, C. Herkommer, A. Billat, D. Grassani, C. Zhang, M. H. P. Pfeiffer, W. Weng, C.-S. Brès, and T. J. Kippenberg, "Mid-infrared frequency comb via coherent dispersive wave generation in silicon nitride nanophotonic waveguides," *Nature Photonics,* vol. 12, no. 6, pp. 330-335, 2018.

[5] Y. Yu, X. Gai, P. Ma, D.-Y. Choi, Z. Yang, R. Wang, S. Debbarma, S. J. Madden, and B. Luther-Davies, "A broadband, quasi-continuous, mid-infrared supercontinuum generated in a chalcogenide glass waveguide," *Laser & Photonics Reviews,* vol. 8, no. 5, pp. 792-798, 2014, 2014.

[6] J. Hult, R. S. Watt, and C. F. Kaminski, "High bandwidth absorption spectroscopy with a dispersed supercontinuum source," *Optics Express,* vol. 15, no. 18, pp. 11385-11395, 2007/09/03/, 2007.

[7] K. J. Kaltenecker, S. Rao D. S, M. Rasmussen, H. B. Lassen, E. J. R. Kelleher, E. Krauss, B. Hecht, N. A. Mortensen, L. Grüner-Nielsen, C. Markos, O. Bang, N. Stenger, and P. U. Jepsen, "Near-infrared nanospectroscopy using a low-noise supercontinuum source," *APL Photonics,* vol. 6, no. 6, pp. 066106, 2021/06/15/, 2021.

[8] L. Wu, Y. Zhang, X. Yuan, F. Zhang, W. Huang, D. Ma, J. Zhao, Y. Wang, Y. Ge, H. Hao, N. Xu, J. Kang, Y. J. Xiang, Y. Zhang, and J. Li, "1D@0D hybrid dimensional heterojunction-based photonics logical gate and isolator," *Applied Materials Today,* vol. 19, 2020/02/28/, 2020.

[9] P. Singh, D. Tripathi, S. Jaiswal, and H. K. Dixit, "All-Optical Logic Gates: Designs, Classification, and Comparison," *Advances in Optical Technologies,* vol. 2014, 2014/03/19/, 2014.

[10] N. Krebs, I. Pugliesi, and E. Riedle, "Pulse Compression of Ultrashort UV Pulses by Self-Phase Modulation in Bulk Material," *Applied Sciences,* vol. 3, no. 1, pp. 153-167, 2013/03//, 2013.

[11] M. F. Marco Peccianti, Luca Razzari, Roberto Morandotti, Brent E. Little, Sai T. Chu, and David J. Moss, "Subpicosecond optical pulse compression via an integrated nonlinear chirper," *Optics Express*, 2010.

[12] Y. Dong, S. Chertopalov, K. Maleski, B. Anasori, L. Hu, S. Bhattacharya, A. M. Rao, Y. Gogotsi, V. N. Mochalin, and R. Podila, "Saturable Absorption in 2D $Ti_3C_2$ MXene Thin Films for Passive Photonic Diodes," *Advanced Materials,* vol. 30, no. 10, pp. 1705714, 2018, 2018.

[13] S. O. Konorov, D. A. Sidorov-Biryukov, I. Bugar, M. J. Bloemer, V. I. Beloglazov, N. B. Skibina, D. Chorvat

Jr, D. Chorvat, M. Scalora, and A. M. Zheltikov, "Experimental demonstration of a photonic-crystal-fiber optical diode," *Applied Physics B,* vol. 78, no. 5, pp. 547-550, 2004/03/01/, 2004.

[14] L. Wu, W. Huang, Y. Wang, J. Zhao, D. Ma, Y. Xiang, J. Li, J. S. Ponraj, S. C. Dhanabalan, and H. Zhang, "2D Tellurium Based High-Performance All-Optical Nonlinear Photonic Devices," *Advanced Functional Materials,* vol. 29, no. 4, pp. 1806346, 2019, 2019.

[15] L. Wu, X. Jiang, J. Zhao, W. Liang, Z. Li, W. Huang, Z. Lin, Y. Wang, F. Zhang, S. Lu, Y. Xiang, S. Xu, J. Li, and H. Zhang, "2D MXene: MXene-Based Nonlinear Optical Information Converter for All-Optical Modulator and Switcher (Laser Photonics Rev. 12(12)/2018)," *Laser & Photonics Reviews,* vol. 12, no. 12, pp. 1870055, 2018, 2018.

[16] S. Moon, and D. Y. Kim, "Ultra-high-speed optical coherence tomography with a stretched pulse supercontinuum source," *Optics Express,* vol. 14, no. 24, pp. 11575-11584, 2006/11/27/, 2006.

[17] Y. S. Norihiko NISHIZAWA, Masahito YAMANAKA, "High Resolution Optical Coherence Tomography Using Ultrashort Pulse Fiber Laser Sources," *ResearchGate*, 2025.

[18] E. Dulkeith, Y. A. Vlasov, X. Chen, N. C. Panoiu, and R. M. Osgood, "Self-phase-modulation in submicron silicon-on-insulator photonic wires," *Optics Express,* vol. 14, no. 12, pp. 5524-5534, 2006/06/12/, 2006.

[19] O. Boyraz, T. Indukuri, and B. Jalali, "Self-phase-modulation induced spectral broadening in silicon waveguides," *Optics Express,* vol. 12, no. 5, pp. 829-834, 2004/03/08/, 2004.

[20] D. T. H. Tan, K. Ikeda, P. C. Sun, and Y. Fainman, "Group velocity dispersion and self phase modulation in silicon nitride waveguides," *Applied Physics Letters,* vol. 96, no. 6, pp. 061101, 2010.

[21] D. Duchesne, M. Ferrera, L. Razzari, R. Morandotti, B. E. Little, S. T. Chu, and D. J. Moss, "Efficient self-phase modulation in low loss, high index doped silica glass integrated waveguides," *Optics Express,* vol. 17, no. 3, pp. 1865-1870, 2009/02/02/, 2009.

[22] W. Bogaerts, and L. Chrostowski, "Silicon Photonics Circuit Design: Methods, Tools and Challenges," *Laser & Photonics Reviews,* vol. 12, no. 4, pp. 1700237, 2018.

[23] S. Feng, T. Lei, H. Chen, H. Cai, X. Luo, and A. W. Poon, "Silicon photonics: from a microresonator perspective," *Laser & Photonics Reviews,* vol. 6, no. 2, pp. 145-177, 2012.

[24] D. J. Moss, *et. al.*, "New CMOS-compatible platforms based on silicon nitride and Hydex for nonlinear optics," *Nat. Photonics,* vol. 7, 2013.

[25] J. Leuthold, C. Koos, and W. Freude, "Nonlinear Silicon Photonics," *Nat. Photonics,* vol. 4, pp. 535-544, 2010.

[26] J. Riemensberger, N. Kuznetsov, J. Liu, J. He, R. N. Wang, and T. J. Kippenberg, "A photonic integrated continuous-travelling-wave parametric amplifier," *Nature,* vol. 612, no. 7938, pp. 56-61, 2022/12//, 2022.

[27] M. Ferrera, L. Razzari, D. Duchesne, R. Morandotti, Z. Yang, M. Liscidini, J. E. Sipe, S. Chu, B. E. Little, and D. J. Moss, "Low-power continuous-wave nonlinear optics in doped silica glass integrated waveguide structures," *Nature Photonics,* vol. 2, no. 12, pp. 737-740, 2008.

[28] M. A. Foster, A. C. Turner, J. E. Sharping, B. S. Schmidt, M. Lipson, and A. L. Gaeta, "Broad-band optical parametric gain on a silicon photonic chip," *Nature,* vol. 441, no. 7096, pp. 960-963, 2006/06//, 2006.

[29] Y. Qu, J. Wu, Y. Zhang, L. Jia, Y. Liang, B. Jia, and D. J. Moss, "Analysis of Four-Wave Mixing in Silicon Nitride Waveguides Integrated With 2D Layered Graphene Oxide Films," *Journal of Lightwave Technology,* vol. 39, no. 9, pp. 2902-2910, 2021/05//, 2021.

[30] Y. Zhang, J. Wu, Y. Qu, L. Jia, B. Jia, and D. J. Moss, "Optimizing the Kerr Nonlinear Optical Performance of Silicon Waveguides Integrated With 2D Graphene Oxide Films," *Journal of Lightwave Technology,* vol. 39, no. 14, pp. 4671-4683, 2021.

[31] T. Gu, N. Petrone, J. F. McMillan, A. van der Zande, M. Yu, G. Q. Lo, D. L. Kwong, J. Hone, and C. W. Wong, "Regenerative oscillation and four-wave mixing in graphene optoelectronics," *Nature Photonics,* vol.

6, no. 8, pp. 554-559, 2012.

[32] A. Ishizawa, R. Kou, T. Goto, T. Tsuchizawa, N. Matsuda, K. Hitachi, T. Nishikawa, K. Yamada, T. Sogawa, and H. Gotoh, "Optical nonlinearity enhancement with graphene-decorated silicon waveguides," *Scientific Reports,* vol. 7, no. 1, pp. 45520, 2017/04/12/, 2017.

[33] N. Vermeulen, D. Castelló-Lurbe, M. Khoder, I. Pasternak, A. Krajewska, T. Ciuk, W. Strupinski, J. Cheng, H. Thienpont, and J. Van Erps, "Graphene's nonlinear-optical physics revealed through exponentially growing self-phase modulation," *Nature Communications,* vol. 9, no. 1, pp. 2675, 2018/07/11/, 2018.

[34] Q. Feng, H. Cong, B. Zhang, W. Wei, Y. Liang, S. Fang, T. Wang, and J. Zhang, "Enhanced optical Kerr nonlinearity of graphene/Si hybrid waveguide," *Applied Physics Letters,* vol. 114, no. 7, pp. 071104, 2019.

[35] P. Demongodin, H. El Dirani, S. Kerdilès, J. Lhuillier, T. Wood, C. Sciancalepore, and C. Monat, "Pulsed Four-Wave Mixing at Telecom Wavelengths in Si3N4 Waveguides Locally Covered by Graphene," *Nanomaterials,* vol. 13, no. 3, pp. 451, 2023.

[36] J. Wu, Y. Yang, Y. Qu, L. Jia, Y. Zhang, X. Xu, S. T. Chu, B. E. Little, R. Morandotti, B. Jia, and D. J. Moss, "2D Layered Graphene Oxide Films Integrated with Micro-Ring Resonators for Enhanced Nonlinear Optics," *Small,* vol. 16, 2020.

[37] Y. Zhang, J. Wu, L. Jia, Y. Qu, Y. Yang, B. Jia, and D. J. Moss, "Graphene Oxide for Nonlinear Integrated Photonics," *Laser & Photonics Rev.,* vol. 17, no. 3, 2023.

[38] Y. Qu, J. Wu, Y. Yang, Y. Zhang, Y. Liang, H. El Dirani, R. Crochemore, P. Demongodin, C. Sciancalepore, C. Grillet, C. Monat, B. Jia, and D. J. Moss, "Enhanced Four-Wave Mixing in Silicon Nitride Waveguides Integrated with 2D Layered Graphene Oxide Films," *Advanced Optical Materials,* vol. 8, no. 23, pp. 2001048, 2020/12/01, 2020.

[39] Y. Zhang, J. Wu, Y. Yang, Y. Qu, L. Jia, T. Moein, B. Jia, and D. J. Moss, "Enhanced Kerr Nonlinearity and Nonlinear Figure of Merit in Silicon Nanowires Integrated with 2D Graphene Oxide Films," *ACS Applied Materials & Interfaces,* vol. 12, no. 29, pp. 33094-33103, 2020/07/22, 2020.

[40] Y. Yang, J. Wu, X. Xu, Y. Liang, S. T. Chu, B. E. Little, R. Morandotti, B. Jia, and D. J. Moss, "Invited Article: Enhanced four-wave mixing in waveguides integrated with graphene oxide," *APL Photonics,* vol. 3, no. 12, pp. 120803, 2018.

[41] Y. Zuo, W. Yu, C. Liu, X. Cheng, R. Qiao, J. Liang, X. Zhou, J. Wang, M. Wu, Y. Zhao, P. Gao, S. Wu, Z. Sun, K. Liu, X. Bai, and Z. Liu, "Optical fibres with embedded two-dimensional materials for ultrahigh nonlinearity," *Nature Nanotechnology,* vol. 15, no. 12, pp. 987-991, 2020/12//, 2020.

[42] A. M. Chiara Trovatello, Xinyi Xu, Changhwan Lee, Fang Liu, Nicola Curreli, Cristian Manzoni, Stefano Dal Conte, Kaiyuan Yao, Alessandro Ciattoni, James Hone, Xiaoyang Zhu, P. James Schuck & Giulio Cerullo, "Optical parametric amplification by monolayer transition metal dichalcogenides," *Nature Photonics*, 2020.

[43] Y. Zhang, L. Tao, D. Yi, J.-B. Xu, and H. K. Tsang, "Enhanced four-wave mixing with MoS2 on a silicon waveguide," *Journal of Optics,* vol. 22, no. 2, pp. 025503, 2020.

[44] I. Alonso Calafell, L. A. Rozema, A. Trenti, J. Bohn, E. J. C. Dias, P. K. Jenke, K. S. Menghrajani, D. Alcaraz Iranzo, F. J. García de Abajo, F. H. L. Koppens, E. Hendry, and P. Walther, "High-Harmonic Generation Enhancement with Graphene Heterostructures," *Advanced Optical Materials,* vol. 10, no. 19, pp. 2200715, 2022, 2022.

[45] L. Liu, K. Xu, X. Wan, J. Xu, C. Y. Wong, and H. K. Tsang, "Enhanced optical Kerr nonlinearity of $MoS_2$ on silicon waveguides," *Photonics Research,* vol. 3, no. 5, pp. 206-209, 2015/10/01/, 2015.

[46] V. P. Yuchen Wang, Samuel Gyger ,Gius Md Uddin, Xueyin Bai,Christian Lafforgue, Laurent Vivien, KlausD. Jöns, Eric Cassan, and Zhipei Sun, "Enhancing Si3N4 Waveguide Nonlinearity with Heterogeneous

Integration of Few-Layer WS2 | ACS Photonics," *ACS Photonics*, 2021.

[47] I. H. Abidi, S. P. Giridhar, J. O. Tollerud, J. Limb, M. Waqar, A. Mazumder, E. L. H. Mayes, B. J. Murdoch, C. Xu, A. Bhoriya, A. Ranjan, T. Ahmed, Y. Li, J. A. Davis, C. L. Bentley, S. P. Russo, E. D. Gaspera, and S. Walia, "Oxygen Driven Defect Engineering of Monolayer MoS$_2$ for Tunable Electronic, Optoelectronic, and Electrochemical Devices," *Adv. Funct. Mater.,* vol. 34, no. 37, 2024.

[48] T. Aung, S. P. Giridhar, I. H. Abidi, T. Ahmed, A. Ai-Hourani, and S. Walia, "Photoactive Monolayer MoS2 for Spiking Neural Networks Enabled Machine Vision Applications," *Advanced Materials Technologies,* vol. 10, no. 16, pp. 2401677, 2025, 2025.

[49] J. Hu, J. Wu, I. H. Abidi, D. Jin, Y. Zhang, J. Mao, A. Pandey, Y. Wang, S. Walia, and D. J. Moss, "Silicon Integrated Photonic Waveguide Polarizers with 2D MoS2 Films," *IEEE Journal of Selected Topics in Quantum Electronics*, pp. 1-12, 2025, 2025.

[50] L. Jia, J. Wu, Y. Zhang, Y. Qu, B. Jia, Z. Chen, and D. J. Moss, "Fabrication Technologies for the On-Chip Integration of 2D Materials," *Small Methods,* vol. 6, no. 3, 2022.

[51] Y. Zhang, J. Wu, Y. Yang, Y. Qu, L. Jia, T. Moein, B. Jia, and D. J. Moss, "Enhanced Kerr Nonlinearity and Nonlinear Figure of Merit in Silicon Nanowires Integrated with 2D Graphene Oxide Films," *ACS Appl. Mater. Interfaces,* vol. 12, no. 29, 2020.

[52] H. Jang, K. P. Dhakal, K.-I. Joo, W. S. Yun, S. M. Shinde, X. Chen, S. M. Jeong, S. W. Lee, Z. Lee, J. Lee, J.-H. Ahn, and H. Kim, "Transient SHG Imaging on Ultrafast Carrier Dynamics of MoS2 Nanosheets," *Advanced Materials (Deerfield Beach, Fla.),* vol. 30, no. 14, pp. e1705190, 2018/04//, 2018.

[53] B. Radisavljevic, A. Radenovic, J. Brivio, V. Giacometti, and A. Kis, "Single-layer MoS2 transistors," *Nature Nanotechnology,* vol. 6, no. 3, pp. 147-150, 2011/03//, 2011.

[54] M. S. Van Luan Nguyen, Junyoung Kwon, Eun-Kyu Lee, Won-Jun Jang, Hyo Won Kim, Ce Liang, Jong Hoon Kang, Jiwoong Park, Min Seok Yoo & Hyeon-Jin Shin, "Wafer-scale integration of transition metal dichalcogenide field-effect transistors using adhesion lithography," *Nature Electronics*, 2023.

[55] S. R. M. Sujay B Desai , Angada B Sachid , Juan Pablo Llinas , Qingxiao Wang , Geun Ho Ahn, Gregory Pitner, Moon J Kim, Jeffrey Bokor , Chenming Hu , H-S Philip Wong , Ali Javey, "MoS2 transistors with 1-nanometer gate lengths," *Science*, 2016.

[56] S. Z. Bowen Li, Jiadong Zhou, Qiao Jiang, Bowen Du, Hangyong Shan, Yang Luo, Zheng Liu, Xing Zhu, Zheyu Fang, "Single-Nanoparticle Plasmonic Electro-optic Modulator Based on MoS2 Monolayers," *ACS Nano*, 2017.

[57] Z. Sun, "Optical modulators with two-dimensional layered materials." pp. 3851-3851.

[58] Y. Zhang, H. Yu, R. Zhang, G. Zhao, H. Zhang, Y. Chen, L. Mei, M. Tonelli, and J. Wang, "Broadband atomic-layer MoS$_2$ optical modulators for ultrafast pulse generations in the visible range," *Optics Letters,* vol. 42, no. 3, pp. 547-550, 2017/02/01/, 2017.

[59] Y. Xie, B. Zhang, S. Wang, D. Wang, A. Wang, Z. Wang, H. Yu, H. Zhang, Y. Chen, M. Zhao, B. Huang, L. Mei, and J. Wang, "Ultrabroadband MoS2 Photodetector with Spectral Response from 445 to 2717 nm," *Advanced Materials,* vol. 29, no. 17, pp. 1605972, 2017, 2017.

[60] O. Lopez-Sanchez, D. Lembke, M. Kayci, A. Radenovic, and A. Kis, "Ultrasensitive photodetectors based on monolayer MoS2," *Nature Nanotechnology,* vol. 8, no. 7, pp. 497-501, 2013/07//, 2013.

[61] D. K. a. G. Konstantatos, "Highly Sensitive, Encapsulated MoS2 Photodetector with Gate Controllable Gain and Speed," *Nano Letters*, 2015.

[62] Y. Qu, J. Wu, Y. Zhang, Y. Yang, L. Jia, H. E. Dirani, S. Kerdiles, C. Sciancalepore, P. Demongodin, C. Grillet, C. Monat, B. Jia, and D. J. Moss, "Integrated optical parametric amplifiers in silicon nitride waveguides incorporated with 2D graphene oxide films," *Light: Advanced Manufacturing,* vol. 4, no. 4, pp. 437, 2023, 2023.

[63] P. Demongodin, H. El Dirani, J. Lhuillier, R. Crochemore, M. Kemiche, T. Wood, S. Callard, P. Rojo-Romeo, C. Sciancalepore, C. Grillet, and C. Monat, "Ultrafast saturable absorption dynamics in hybrid graphene/Si$_3$N$_4$ waveguides," *APL Photonics,* vol. 4, no. 7, 2019.

[64] H. El Dirani, A. Kamel, M. Casale, S. Kerdiles, C. Monat, X. Letartre, M. Pu, L. K. Oxenløwe, K. Yvind, and C. Sciancalepore, "Annealing-free Si3N4 frequency combs for monolithic integration with Si photonics," *Applied Physics Letters,* vol. 113, no. 8, pp. 081102, 2018.

[65] Y. Y. Alper Gurarslan, Liqin Su, Yiling Yu, Francisco Suarez, Shanshan Yao, Yong Zhu, Mehmet Ozturk, Yong Zhang, Linyou Cao, "Surface-Energy-Assisted Perfect Transfer of Centimeter-Scale Monolayer and Few-Layer MoS2 Films onto Arbitrary Substrates," *ACS Nano,* 2014.

[66] I. H. Abidi, A. Bhoriya, P. Vashishtha, S. P. Giridhar, E. L. H. Mayes, M. Sehrawat, A. K. Verma, V. Aggarwal, T. Gupta, H. K. Singh, T. Ahmed, N. D. Sharma, and S. Walia, "Oxidation-induced modulation of photoresponsivity in monolayer MoS2 with sulfur vacancies," *Nanoscale,* vol. 16, no. 42, pp. 19834-19843, 2024/10/31/, 2024.

[67] I. H. Abidi, L. T. Weng, C. P. J. Wong, A. Tyagi, L. Gan, Y. Ding, M. Li, Z. Gao, R. Xue, M. D. Hossain, M. Zhuang, X. Ou, and Z. Luo, "New Approach to Unveiling Individual Atomic Layers of 2D Materials and Their Heterostructures," *Chemistry of Materials,* vol. 30, no. 5, pp. 1718-1728, 2018/03/13/, 2018.

[68] C. Koos, P. Vorreau, T. Vallaitis, P. Dumon, W. Bogaerts, R. Baets, B. Esembeson, I. Biaggio, T. Michinobu, F. Diederich, W. Freude, and J. Leuthold, "All-optical high-speed signal processing with silicon-organic hybrid slot waveguides," *Nature Photonics,* vol. 3, pp. 216-219, 2009/04/01/, 2009.

[69] H. Arianfard, S. Juodkazis, D. J. Moss, and J. Wu, "Sagnac interference in integrated photonics," *Applied Physics Reviews,* vol. 10, no. 1, pp. 011309, 2023/03/01, 2023.

[70] H. Arianfard, J. Wu, S. Juodkazis, and D. J. Moss, "Optical Analogs of Rabi Splitting in Integrated Waveguide-Coupled Resonators," *Advanced Physics Research,* vol. 2, no. 9, pp. 2200123, 2023, 2023.

[71] Y. Zhang, J. Wu, Y. Yang, Y. Qu, H. E. Dirani, R. Crochemore, C. Sciancalepore, P. Demongodin, C. Grillet, C. Monat, B. Jia, and D. J. Moss, "Enhanced Self-Phase Modulation in Silicon Nitride Waveguides Integrated With 2D Graphene Oxide Films," *IEEE Journal of Selected Topics in Quantum Electronics,* vol. 29, no. 1: Nonlinear Integrated Photonics, pp. 1-13, 2023/01//, 2023.

[72] Y. Zhang, J. Wu, Y. Yang, Y. Qu, L. Jia, H. E. Dirani, S. Kerdiles, C. Sciancalepore, P. Demongodin, C. Grillet, C. Monat, B. Jia, and D. J. Moss, "Enhanced Supercontinuum Generation in Integrated Waveguides Incorporated with Graphene Oxide Films," *Advanced Materials Technologies,* vol. 8, no. 9, pp. 2201796, 2023/05//, 2023.

[73] O. A. S. Koen Alexander, Sergey A. Mikhailov, OrcidBart Kuyken, Dries Van Thourhout, "Electrically Tunable Optical Nonlinearities in Graphene-Covered SiN Waveguides Characterized by Four-Wave Mixing," *ACS Photonics*, 2017.

[74] J. Hu, J. Wu, W. Liu, D. Jin, H. E. Dirani, S. Kerdiles, C. Sciancalepore, P. Demongodin, C. Grillet, C. Monat, D. Huang, B. Jia, and D. J. Moss, "2D Graphene Oxide: A Versatile Thermo-Optic Material," *Advanced Functional Materials,* vol. n/a, no. n/a, pp. 2406799.

[75] C.-H. Wu, T. Reep, Y. Huang, S. Brems, C. Haffner, C. Huyghebaert, B. Kuyken, J. Van Campenhout, M. Pantouvaki, D. Van Thourhout, and Y. Guo, "Low-loss phase modulation using a MoS2 monolayer integrated on silicon waveguides," *2D Materials,* vol. 12, no. 4, pp. 045006, 2025.

[76] G. P. Agrawal, "Nonlinear Fiber Optics," *Nonlinear Science at the Dawn of the 21st Century*, P. L. Christiansen, M. P. Sørensen and A. C. Scott, eds., pp. 195-211, Berlin, Heidelberg: Springer Berlin Heidelberg, 2000.

[77] C. Donnelly, and D. T. H. Tan, "Ultra-large nonlinear parameter in graphene-silicon waveguide structures," *Optics Express,* vol. 22, no. 19, pp. 22820-22830, 2014/09/22/, 2014.

[78] M. Ji, H. Cai, L. Deng, Y. Huang, Q. Huang, J. Xia, Z. Li, J. Yu, and Y. Wang, "Enhanced parametric frequency conversion in a compact silicon-graphene microring resonator," *Optics Express,* vol. 23, no. 14, pp. 18679-18685, 2015/07/13/, 2015.